\documentclass[letterpaper, 10 pt, conference]{ieeeconf} 

\IEEEoverridecommandlockouts
\usepackage[hyperindex,hidelinks]{hyperref}
\hypersetup{
  pdfinfo     = {
    Title    = {Warmth and Competence to Predict Human Preference of Robot Behavior in Physical Human-Robot Interaction},
    Author   = {Marcus M. Scheunemann and Raymond H. Cuijpers and Christoph Salge},
    Subject  = {A solid methodology to understand human perception and preferences in human-robot interaction (HRI) is crucial in designing real-world HRI. Social cognition posits that the dimensions Warmth and Competence are central and universal dimensions characterizing other humans. The Robotic Social Attribute Scale (RoSAS) proposes items for those dimensions suitable for HRI and validated them in a visual observation study. In this paper we complement the validation by showing the usability of these dimensions in a behavior based, physical HRI study with a fully autonomous robot. We compare the findings with the popular Godspeed dimensions Animacy, Anthropomorphism, Likeability, Perceived Intelligence and Perceived Safety. We found that Warmth and Competence, among all RoSAS and Godspeed dimensions, are the most important predictors for human preferences between different robot behaviors. This predictive power holds even when there is no clear consensus preference or significant factor difference between conditions.},
    Lang     = {en-US}
  }
}

\usepackage{graphics}
\usepackage{amsmath}
\usepackage{amssymb} 
\usepackage{bm}

\usepackage{cite}

\usepackage[shortcuts,acronym,nohypertypes={acronym}]{glossaries}
\newacronym{HRI}{HRI}{human-robot interaction}
\newacronym{RoSAS}{RoSAS}{Robotic Social Attribute Scale}

\usepackage{siunitx}
\usepackage{xcolor}
\definecolor{color1}{RGB}{44,123,182}
\definecolor{color2}{RGB}{252,141,89}
\definecolor{color3}{RGB}{145,191,219}
\usepackage{booktabs}
\usepackage{subcaption}
\usepackage[disable]{todonotes}
\usepackage{multirow,bigdelim}

\title{\LARGE \bf
Warmth and Competence to Predict Human Preference of Robot Behavior in Physical Human-Robot Interaction}

\author{Marcus M. Scheunemann$^{1}$ and Raymond H. Cuijpers$^{2}$ and Christoph Salge$^{1}$%
\thanks{$^{1}$Marcus M. Scheunemann and  Christoph Salge are with the School of Computer Science,
        University of Hertfordshire, AL10 0UT, Hatfield, UK
        {\tt\small marcus@mms.ai} and {\tt\small christophsalge@gmail.com}}%
\thanks{$^{2}$Raymond H. Cuijpers,
        Eindhoven University of Technology, 5600 MB, Eindhoven, The Netherlands
        {\tt\small r.h.cuijpers@tue.nl}}%
}

\begin{document}

\maketitle
\thispagestyle{empty}
\pagestyle{empty}

\begin{abstract}
A solid methodology to understand human perception and preferences in human-robot
interaction~(HRI) is crucial in designing real-world HRI.
Social cognition posits that the dimensions Warmth and Competence are central and universal dimensions characterizing other humans~\cite{FiskeCuddyEtAl-07}.
The Robotic Social Attribute Scale~(RoSAS) proposes items for those dimensions suitable for HRI and validated them in a visual observation study.
In this paper we complement the validation by showing the usability of these dimensions in a behavior based, physical HRI study with a fully autonomous robot. We compare the findings with the popular Godspeed dimensions Animacy, Anthropomorphism, Likeability, Perceived Intelligence and Perceived Safety.
We found that Warmth and Competence, among all RoSAS and Godspeed dimensions, are the most important predictors for human preferences between different robot behaviors. This predictive power holds even when there is no clear consensus preference or significant factor difference between conditions. 
\end{abstract}

\section{INTRODUCTION}
There is a large body of work evaluating the perception of and interaction with robots. In this paper we are interested in understanding which metrics indicate human preferences, i.e., which robot a person would choose to interact with again, if given a choice. Agreeing upon a metric for this in \ac{HRI} would provide important benefits~\cite{steinfeld2006common}, but raises the question which metric we should use? The human engagement in an interaction could serve as an indicator for their preference. However, measuring engagement is a time consuming task and compatibility between results is difficult due to a lack of a standardized coding strategy.
A common, alternative approach for evaluating human-robot interactions is the use of questionnaires. 
Bartneck et al. designed the Godspeed scale~\cite{BartneckKulicEtAl-09}, which captures five dimensions: Anthropomorphism, Animacy, Likeability, Perceived Intelligence and Perceived Safety. It found many uses in the \ac{HRI} community~\cite{WeissBartneck-15}, such as the evaluation of new robot designs. 

In general, it would be beneficial to have both standardized and well validated metrics \cite{steinfeld2006common,WeissBartneck-15}. However, there are some concerns about the Godspeed questionnaire design~\cite{HO2010}. One is that the items encompassing the Animacy and Anthropomorphism dimensions load onto each other. In fact, they even share an item making them overlap by design~\cite{WeissBartneck-15}.
Another critique is the use of the semantic differential scale~\cite{BartneckKulicEtAl-09}. While, e.g., \emph{Unfriendly--Friendly} are clear antonyms, some differentials, such as \emph{Machinelike--Humanlike}, are not necessarily entities of a bipolar scale. This makes the items challenging to answer at times. %
In \cite{CarpinellaWymanEtAl-17} a factor analysis was used to validate the dimensions of the Godspeed questionnaire using a large pool of participants ($N=215$). It was shown that the Godspeed questionnaire items for Likeability and Perceived Intelligence mostly loaded highly on independent factors, but several questionnaire items for Anthropomorphism loaded more strongly on the factor for Animacy and vice versa (cf.~Table~1 in \cite{CarpinellaWymanEtAl-17}). This suggests that the Animacy and Anthropomorphism dimensions are not reliable. For Perceived Safety, only two out of three items loaded on an independent factor, while the third loaded most strongly on the factor corresponding to Animacy. But the main question here is how well do these five factors correspond to human preferences for interactive robot behaviors?

In cognitive science and social psychology Warmth and Competence are considered fundamental dimensions of social cognition, i.e., the social judgment of our peers~\cite{FiskeCuddyEtAl-07,JuddJames-HawkinsEtAl-05}.
Fiske et al. provide evidence that those dimensions are universal and reliable for social judgment across stimuli, cultures and time~\cite{FiskeCuddyEtAl-07}.
People perceived as warm and competent elicit uniformly positive emotions~\cite{FiskeCuddyEtAl-07}, are in general more favored, and experience more positive interaction with their peers~\cite{CarpinellaWymanEtAl-17}. The opposite is true for people scoring low on these dimensions, meaning they experience more negative interactions~\cite{FiskeCuddyEtAl-07}.
Warmth and Competence, together, almost entirely account for how people perceive and characterize others~\cite{FiskeCuddyEtAl-07}, making them main drivers for how humans judge one another.
There are different classifications for people scoring high on one dimension only, but they are similarly socially important.
For example, people scoring high on Warmth but low on Competence elicit sympathy or pity~\cite{JuddJames-HawkinsEtAl-05,CuddyFiskeEtAl-07}, while those scoring high on both elicit admiration.
The Warmth dimension, however, carries more weight in inter persona judgments, like affect and behavioral reactions~\cite{FiskeCuddyEtAl-07}.

Items for the dimensions Warmth and Competence are proposed for \ac{HRI} studies in the \ac{RoSAS}~\cite{CarpinellaWymanEtAl-17,Stroessner-20}.
The authors derived the dimension items and validated them in four studies with the use of robot images. For example, one study reproduced an established human judgment tied to gender, namely that females are perceived as more warm than males.
This stimulus was investigated and evidence confirms that robots appearing more feminine were perceived as more warm than robots appearing more masculine.

Along the creation and validation process in \cite{CarpinellaWymanEtAl-17} an additional dimension, Discomfort, emerged. According to \cite{CarpinellaWymanEtAl-17}, this factor does not appear in measures of social perception of humans, but provides an additional and robust scale in the evaluation of robots. It was included in \ac{RoSAS} because \ac{HRI} researchers may be interested in discomfort elicited by their robot.
Each of the three dimensions computes from a set of 6 items, and 
a factor analysis presented in Table~3 of \cite{CarpinellaWymanEtAl-17} shows that the items load on the dimensions presented.
This is indeed a promising outcome. A relatively short and easily applicable questionnaire offering at least two dimensions explaining our full perception of robots has the potential to simplify and streamline \ac{HRI} studies and drive the field forward.

However, it is not yet fully understood whether those dimensions are indeed embodied~\cite{GoldmanVignemont-09}, i.e., persist beyond preference evaluation based on images. Evidence suggests that humans react differently to a real embodied robot. The simple act of moving can already change the perception of a robot~\cite{Dautenhahn-97}. There is ample evidence that embodiment plays an important role in social cognition~\cite{GoldmanVignemont-09}, and hence is important for \ac{HRI} as well.
While robot images may help us to infer social perception of a robot and create a suitable hardware, only an embodied and moving robot will help us to fully understand the traits and perception elicit by a robot. The main aim of this paper is to investigate if the dimensions Warmth and Competence are useful for physical, embodied interaction studies.

A popular approach would be to use a factor analysis for validation. A notable disadvantage however is the large number of participants needed for this approach, making it particularly unsuitable for a physical interaction study. For example, study 2 for validating the \ac{RoSAS} had 209 participants~\cite{CarpinellaWymanEtAl-17}. 
A classical scale validation among several \ac{HRI} studies with various robot platforms is a very complex and time consuming task. This paper aims to close this gap by suggesting a paradigm for gathering evidence for the usability of dimensions for physical \ac{HRI} studies, thus complementing the original validation of~\cite{CarpinellaWymanEtAl-17}.
Our paradigm does not require a factor analysis but uses Bayesian analysis. We relate self-reported human interaction preference with preference estimations derived from the questionnaire dimensions.

\subsection{Research Question}
Are Warmth and Competence the best predictors for human interaction preferences among all dimensions of the \ac{RoSAS} and Godspeed scale?
Or, in other words, do they indicate that one particular robot behavior is favored by humans, the same way they indicate this for human-human interaction in social cognition?

\subsection{Overview}
The perception of Warmth and Competence has been validated
in~\cite{CarpinellaWymanEtAl-17} in a study based on human visual
perception of robotic still images. In contrast, our study focuses on robot
behavior. In \autoref{sec:study_design} we present the study design.
Subsection~\ref{sec:robot_environment} presents the environment and the
robot platform: a minimal, but fully autonomous robot (based on taxonomy
in~\cite{Beer:2014}). By fully autonomous we mean the robot's behavior is
neither remote controlled by a human, nor scripted previously, but is
instead generated by an algorithm that reacts to external sensor stimuli.
All participants interact with the same robot platform, which is minimal in
that it has few degrees of freedom and is not a humanoid.
Participants are exposed to three different robot behaviors.
We decided for autonomous behavior generation algorithms, which we set up so that the robot behaviors in all conditions appear very similar~(cf.~\autoref{sec:conditions} or the supplementary video~\cite{supp}).
This resulted in no clear consensus preference or significant factor difference between conditions. Instead, participants' responses are mainly based on their interaction experience.
The \autoref{sec:procedure} describes how we collect the responses for all
dimensions of the \ac{RoSAS} and Godspeed scale. The \autoref{sec:variables} describes the evaluation of all dimensions as predictors for the participant's self-reported preferred interaction. That way, we can also compare the strength of Warmth and Competence to other popular dimensions used in \ac{HRI}.

Section~\ref{sec:results} presents the results, which are discussed in \autoref{sec:discussion}. 
The result show (i) only small dimension difference between conditions and (ii) no clear overall condition preference.
This was intended and a central objective of the study design. It minimizes the chance of a common cause explanation, which would cause an arbitrary influence on the dimensions.
This way, the results for a correspondence between the participant's self-reported preference and the prediction of their preference using the dimensions has more weight.
Our main results show that Warmth and Competence are indeed the most important dimensions for predicting participants' interaction preferences. 
This indicates that, similar to inter persona interaction in social cognition, we prefer to interact with robots perceived as more warm~\cite{FiskeCuddyEtAl-07,JuddJames-HawkinsEtAl-05}.

\section{STUDY DESIGN}
\label{sec:study_design}
The following subsections describe the design of the study. Most importantly, we aim for three conditions with the same range of behavior patterns.
The robot used has only two degrees of freedom, but generates its behavior autonomously in a tight feedback loop able to react quickly to external stimuli. The human participant is encouraged to physically interact with the robot. That way, we investigate the usability of questionnaire dimensions on an interaction level rather than a visual level, as done by the authors of \ac{RoSAS}~\cite{CarpinellaWymanEtAl-17}.

\subsection{Robot \& Environment}
\label{sec:robot_environment}
\begin{figure}[htbp]
  \centering
  \includegraphics[width=\linewidth]{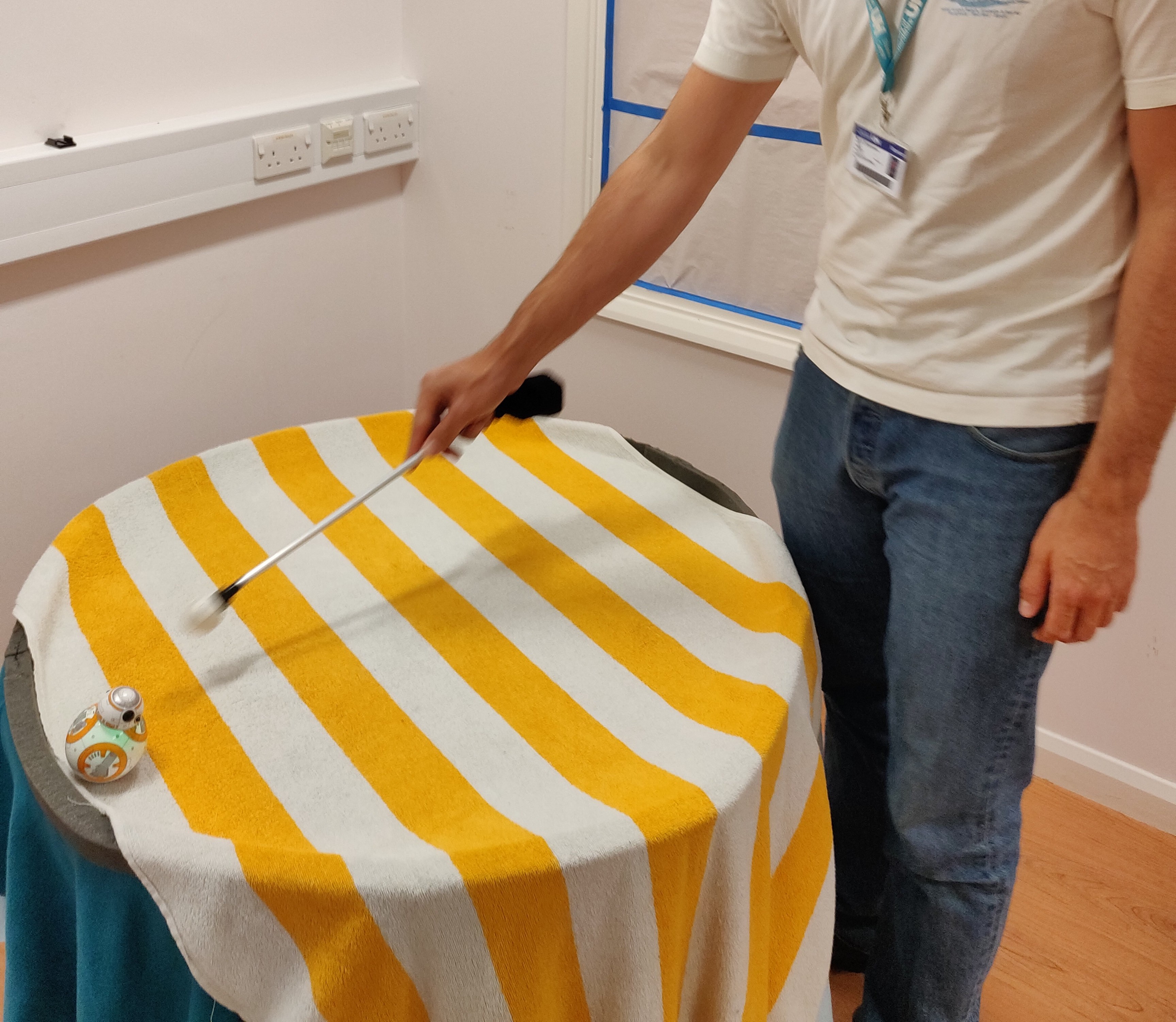}
  \caption{The experimental environment showing the robot platform Sphero in its BB8 version and a participant using a tool to interact with the robot. The robot can freely locomote on the table. The participant can move around the table for observing or interacting with the robot.}
  \label{fig:environment_robot}
\end{figure}
\autoref{fig:environment_robot} shows the robot and the experimental environment of the study.
We use the spherical robot Sphero in its BB8 version~\cite{ScheunemannSalgeEtAl-19,sphero2020}. The head is attached to the robot's inner vehicle with a magnet, providing the participant with a sense of the robot's direction.
The robot has only two degrees of freedom, controlling the wheel speed of the two wheels attached to the inner vehicle. 
However, the behavior patterns are diverse. The robot can spin, turn, move straight or wobble.
The behavior generation is described in \autoref{sec:conditions}.

\autoref{fig:environment_robot} also shows the environment the robot can freely locomote in. The table is \SI{91}{\centi\meter} in diameter.
We provided a wand-shaped tool to motivate interaction with the robot.
Participants were allowed to choose any position around the table and could change their position at will, described in \autoref{sec:procedure}.

\subsection{Conditions (Robot Behavior Differences)}
\label{sec:conditions}
The study consists of three conditions (labeled A, B and C) with the same robot platform, but with a slightly different robot behavior per condition.
The idea was to have behaviors which are very similar to each other, so participants do not understand the purpose of the study.
A video supplementing this submission shows an example of all three conditions conducted by one, randomly chosen participant~\cite{supp}.
The remainder of the section describes how the behavior is generated.

The controllers generating the robot behavior differ between conditions by the sensors used as input or by the update rules of the controller parameter. \autoref{tab:controller-difference} provides an overview of the differences as per condition.

\begin{table}[tbh]
    \centering
    \caption{Overview of the 3 experimental conditions}
    \begin{tabular}{cll}
        \toprule
        Condition & Sensor Input & Network Update \\
        \midrule
        A & no proximity sensor & online adaptation with PI\\
        B & proximity sensor & based on replay\\
        C & proximity sensor & online adaptation with PI\\
        \bottomrule
    \end{tabular}
    \label{tab:controller-difference}
\end{table}

\paragraph{Sensor Input}
All controllers receive readings from an accelerometer, a gyroscope and the servos. The gyroscope provides the angular velocity around the central axis from head to bottom shell and the accelerometer provides the linear acceleration along the forward and sideward axes. Each of the two servos provides its current speed.
In condition B and C, the controller has an additional input: a one-dimensional proximity sensor corresponding to the distance of the interaction wand. %
The proximity information is derived from the signal strength between two Bluetooth Low Energy devices~\cite{ScheunemannDautenhahnEtAl-16}. 
This way the robot can \emph{distinguish} between perturbations by the environment or by the participant. %

\paragraph{Updating Network Weights}
In condition A and C, the robot was equipped with a computational model of intrinsic motivation~\cite{oudeyer2009intrinsic}. The update rules for the network are implemented by time-local predictive information~\cite{martius2013information,der2012playful,ScheunemannSalgeEtAl-19}.
The robot tries to excite different sensors through the generation of a variety of motion regimes, but in a predictable way.
For example, the robot may spin around to excite the gyroscope, or accelerate to excite the forward acceleration measured by the accelerometer.
The implementation and a more detailed description can be found in~\cite{martius2013information,ScheunemannSalgeEtAl-20}.
In condition B, the robot controller is not updated by predictive information, but by replaying network weight updates of an earlier run with a predictive information controller. This means, it changes its network weights, but it is not adaptive toward the current environment or the current participant.

This means the robot is reactive toward the sensory input in all conditions. However, the update of the network weights happens either by maximizing predictive information (A and C), or by replaying (B) the adaptation that happened in a different experiment. 
Overall, the regimes of generated behaviors are very similar, alas not adaptive toward the environment in B. This similarity in behavior will be reflected in similarly perceived factors. It will be shown that the main effects for all dimensions are indeed very similar, i.e., comparing the mean difference of the dimensions between each condition is not statistically significant, with only a few exceptions~(see~\autoref{sec:results}).

\subsection{Procedure}
\label{sec:procedure}
In every session, each participant conducted three interactions with the same robot platform.
The behavior generation is a dependent within-subject variable, i.e., all participants were exposed to all three conditions A, B and C.
The order of the conditions was randomly assigned and counterbalanced. %
Participants were provided with only one task: to find out if the robot behaves differently in each condition.
For solving that task, participants could use a wand to interact with the robot. They received an introduction on how to nudge and push the robot before starting.
They were not provided with any further information on the robot platform, its behavior, or any further description.

After each condition, the participants responded to a questionnaire, encompassing all \ac{RoSAS} dimensions\footnote{RoSAS dimensions: Warmth, Competence and Discomfort} on 7-point Likert scales and the Godspeed dimensions\footnote{Godspeed dimensions: Anthropomorphism, Animacy, Likeability, Perceived Intelligence and Perceived Safety} on 5-point semantic differential scales.

After all three conditions had been presented, i.e., at the end of the session, participants were asked about their preferred interaction with the question: ``If you could interact with one of the robots again, which one would you choose?". They could answer with the number of their preferred interaction 1, 2 or 3. 
They could also tick ``no preference''.

\subsection{Participants}
\label{sec:sample}
The sample consists of 36 participants (11 female, 24 male and one who wished not to further specify). They are between 19 to 62 years old ($M=33.56,\ SD=10.24
$). The participants' background is mostly computer science. However, 14 participants have no background in computer science or related fields. 12 participants are not associated to the university where the study is conducted.
All participants are na\"ive toward the study idea and the research interest of the experimenter.

\subsection{Variables}%
\label{sec:variables}
Let $\mathcal{C}$ be the set of all three conditions and $\mathcal{D}$ be the 
set of all scale dimensions.
Then there are three variables dependent on the participant's responses:
\begin{align}
  o &\in\mathcal{C}, \text{ observed preferred condition (self reported)} \nonumber\\  
  r_{d,c} &\in\mathbb{R}, \text{ scale response to $d\in\mathcal{D}$ in $c\in\mathcal{C}$} \nonumber\\
  e &\in\mathcal{C}, \text{ expected preferred condition}\nonumber
\end{align}\todo{$\mathbb{R}$ replace with [1,7] and [1,5]?}%
The \emph{observed preferred condition} $o$ is retrieved directly from the 
participant's answer to the question about their preference. 
The scale response $r$ is computed from the questionnaire responses.

The \emph{expected preferred condition} $e$ is the condition which returns 
the highest participant's scale response value for a specific dimension.
More formally, let $\hat{d}\in\mathcal{D}$ be the dependent scale dimension and 
let $\bm{R}_{\hat{d}} = \{r_{\hat{d},A},\ r_{\hat{d},B},\ 
r_{\hat{d},C}\}$ be a sequence of all three scale responses to the predictor
dimension $\hat{d}$, then:

\begin{equation}
e_{\hat{d}} = c,\ \text{if } r_{c,\hat{d}} = \max_{c}{\{\bm{R}_{\hat{d}}\}}\ 
\text{and } r_{c,\hat{d}} \neq r_{c,d} \forall d\in\mathcal{D}\setminus\hat{d}
\label{eq:expected_preferred}
\end{equation}

An example: let $\hat{d}=\text{Warmth}$ be the dimension which is used as 
discriminator to predict the participant's preference $e$.
The responses of one participant to the three conditions are given as 
$\bm{R}_{\text{Warmth}} = \{3,\ 4.3,\ 2\}$. 
Then $e_\text{Warmth}=B$, since $r_{B,\text{Warmth}} = \max\{\bm{R}_{\text{Warmth}}\} = 4.3$.

Furthermore, it seems sensible to assume that the condition with the lowest scoring for Discomfort could serve as a valid predictor for participants' most preferred interaction.
We therefore decided to extend the set $\mathcal{D}$ with the additional predictor named Discomfort\textsuperscript{--}, which is the inverted dimension of Discomfort. This means \autoref{eq:expected_preferred} computes the expected preferred condition $e$ based on the lowest response value of Discomfort.

\subsection{Data Preparation}
After conducting the study both standardized questionnaires and their dimensions are analyzed for their usability.
Cronbach's $\alpha$ is used to test for internal consistency reliability of the scale dimensions. The item \emph{Quiescent--Surprised} loaded negatively on the Godspeed factor Perceived Safety and was removed.
All dimensions show good reliability~($0.79<\alpha{}<0.92$). 
After that, an analysis of variances\footnote{Computed with \texttt{aov(dimension {\raise.17ex\hbox{$\scriptstyle\mathtt{\sim}$}} condition * order)}, a function part of base \texttt{R}'s in-built \texttt{stats} package.} of the questionnaire responses was conducted. The results show that the condition and their order are not interacting for any of the scale response variables, which allows us to analyze all questionnaire dimensions independently of their order.

\subsection{Data Analysis}
Within-subjects designs are common in \ac{HRI}. 
If interaction effects between conditions and their order can be ruled out,
the main effects can be analyzed with a pairwise comparison of condition responses.
Pairwise tests analyze the \emph{change} of participants' answers, rather than
comparing \emph{all} answers of one experimental group to another group.

In our study, participants are not given any context but the question to explore whether the robots in the conditions are different. 
This is to avoid framing participants' expectations, but rather leaving the participants alone to their own experiences.
This however influenced the choice of data analysis, since we expected responses at both end of the scale.
For example, a person who expected the robot to behave and speak like the Star Wars character might be disappointed by the robot's behavior and mainly respond to dimensions on the lower end of the scale.
In contrast, a person without too many expectations might be excited about the robot's behavior and always answer on the opposite side, and so on.
Averaging their answers of one condition and comparing them to another
(non-pairwise test) will therefore be less conclusive than comparing
whether participants usually rate the one condition higher than the other
(pairwise test).

Therefore, we hypothesize that investigating how all responses to $o$ are dependent
on the response value $r$ does not yield much information in our setting.
This is why we do not conduct a regression analysis, such as logistic regression or
an analysis of variances~(ANOVA).
Instead, \autoref{eq:expected_preferred}
allows us to analyze \emph{if} and \emph{how} the condition with the highest response
to a dimension ($e$) is associated to the participant's reported preferred condition ($o$).

The analysis which answers this consists of two parts: firstly a dependency analysis, which will show if there is \emph{any} dependent association between $o$ and $e$.
Secondly, a correspondence analysis which will show how the levels of $o$ and $e$ are corresponding, i.e., do participants tend to report to like C, while also responding highest to the dimension Warmth? More formally: $e_\text{Warmth} = C = o$? 

\section{RESULTS}%
\label{sec:results}%
\subsection{Main Effects}
\begin{figure}[!htbp]
  \includegraphics{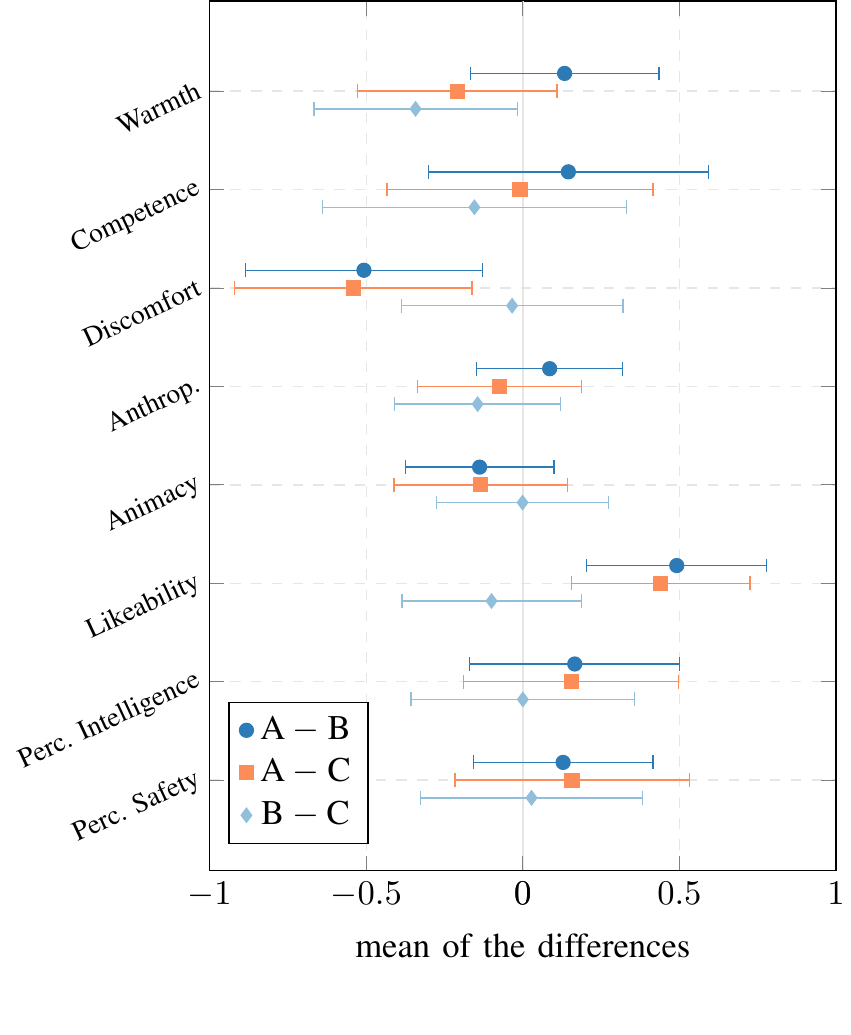}
  \caption{Paired $t$-Test results for all condition combinations. Depicted is the mean of the differences and the 95\% confidence interval as error bars.
  For example, a positive mean of difference for the condition pair \textcolor{color2}{$\text{A}-\text{C}$} indicates that the responses to, e.g., Warmth, are higher for A than for C.
}
\label{fig:main-effects}
\end{figure}%
\autoref{fig:main-effects} shows the main effects of the questionnaire dimensions computed with a pairwise $t$-Test. It can be seen that the largest effects are comprised by Likeability and Discomfort. The mean of the differences is statistically significantly higher for A compared to the other conditions B and C (and the other way around for Discomfort). Other than that, most effects are very small, i.e., point estimates close to zero and large error bars.

\begin{table}[htbp]
  \centering
  \caption{Reported preferred conditions $o$.}
  \begin{tabular}{cccc}
    \toprule
    A & B & C & none\\
    \midrule
    11 & 7 & 13 & 5\\
    \bottomrule
  \end{tabular}  
  \label{tab:preferences}
\end{table}

\autoref{tab:preferences} shows the participant's responses to the question which collects the participants' preferred condition $o$. It can be seen that there is no statistically significant preference for any of the conditions. %

Overall, the pairwise main effects depicted in \autoref{fig:main-effects} are small and there is no statistically significant preferred condition $o$. This has been part of our requirement for the underlying investigation. With only small or no effects, we do not bias the participants toward one condition.
For example, if we had chosen a non-moving and a moving robot for the conditions, the preference of participants would likely be the moving robot with a high response value for many dimensions. This minimizes the chance for a common cause explanation, i.e., that there is an element in one of the conditions that would both cause participants to prefer a given condition and also make them rate that condition highly in a given dimension. 
With our approach, we blur the perception and thus concentrate on the non-obvious perceptions of the robot.

\subsection{Qualitative Observations}
\begin{figure}[htbp]
  \centering
  \includegraphics[width=\columnwidth]{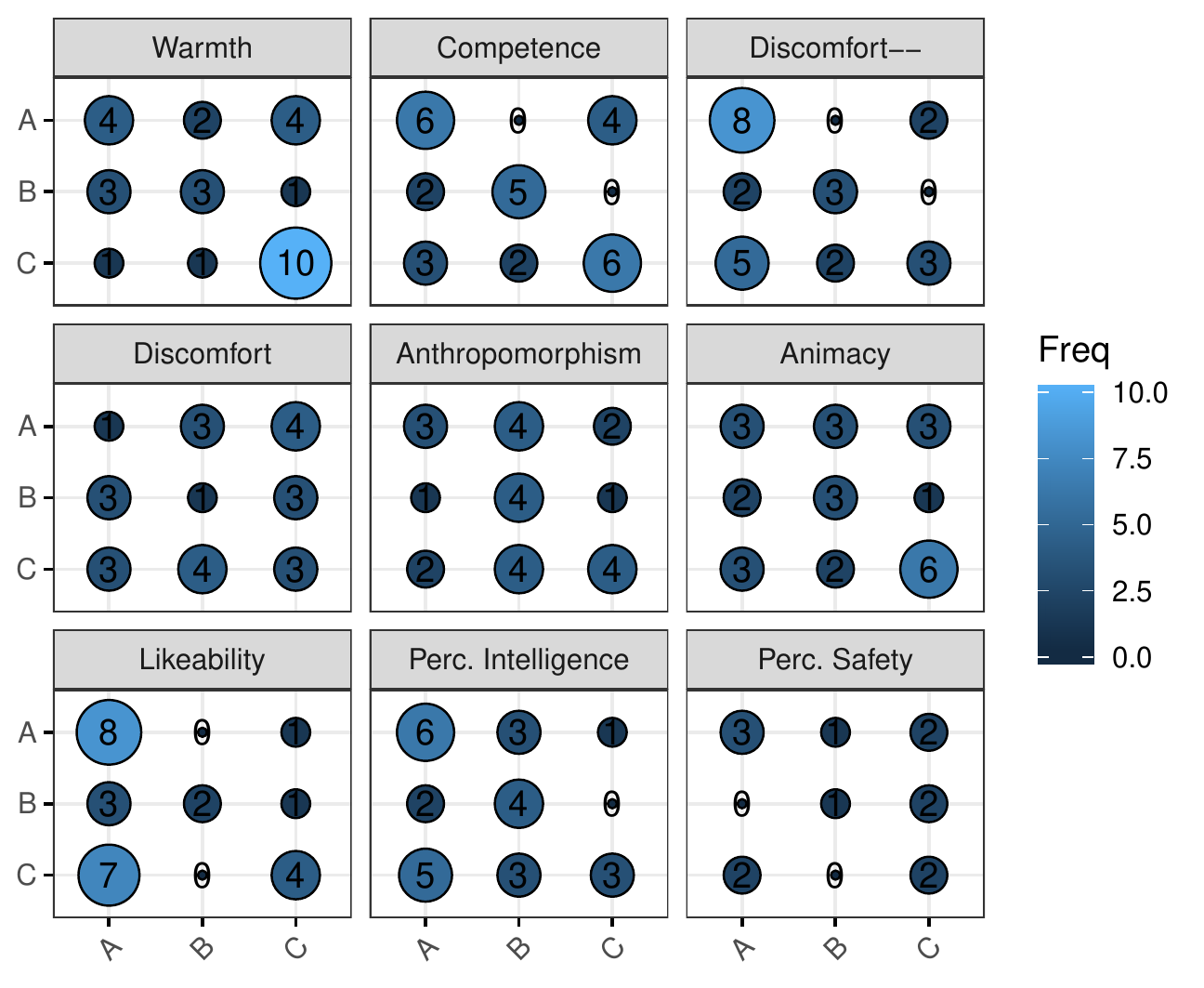}
  \caption{The contingency tables of each dimension $d$ for the observed 
    preferred condition $o$ (rows) and the expected preferred condition $e_d$ 
    (columns) as balloon plots. 
    The larger the size and the lighter the blue, the higher is the cell 
    frequency.}
  \label{fig:results:balloon_plot}
\end{figure}

\autoref{fig:results:balloon_plot} gives a first impression of the data 
frequencies with contingency tables of the central two variables $e$~and~$o$.
The balloon plots are a visualization of a contingency table for 
each of the dimensions $d$ of the questionnaires.
The rows are the levels of the observed preferred condition~$o$, the columns 
are the levels of the expected preferred condition~$e$.
A perfect correspondence of $e$ and $o$ would result in high frequencies on the 
main diagonal and zeros elsewhere.
It can be seen that for the dimensions Warmth and Competence the frequencies along
the main diagonal are high. We learned that Warmth has a small, but statistically significant effect for C compared to B.
This seems to be reflected in \autoref{fig:results:balloon_plot}: it confirms that participants do prefer condition C, and, even more important, that this condition is also associated with high perceived Warmth.
On the other hand, there are no main effects for the Competence dimension,
which is likewise reflected in the figure, as there is no unique high frequency.
However, it can be seen that there is an almost equal distribution of 
frequencies along the main diagonal. This means, despite the absence of a main 
effect, the participants' preference seems to correspond to their responses to 
the Competence dimension.

It occurs that the conditions with the highest responses to the dimensions Warmth and 
Competence do correspond to the participant's reported preferences.
There is no similar observation for the other dimensions. Somewhat counter intuitively, Likeability and Discomfort\textsuperscript{--} do not show such an association.
The remainder of this section will quantify these observation.

\subsection{Dependency}
The dependency analysis uses the Fisher's exact test\footnote{The Fisher's exact test is the exact version
of the popular Chi-squared Test, which cannot be used here because the requirements for expected cell values
are not met}.
It tests the null hypothesis that the two categorical variables $e$ and $o$ are independent.

\begin{table}[th]
    \centering
    \caption{Fisher's exact test and uncertainty coefficient.}
  \begin{tabular}{lccc}
    \toprule
    $d$ (predictor) & $p$ value & $N$ & $U(o|e)$\\\midrule
    Warmth     & .033 & 29 & .178\\
    Competence & .007 & 28 & .135\\ 
    Discomfort & .669 & 25 & .054\\
    Discomfort\textsuperscript{--} & .091 & 25 & .08\\\midrule
    Anthropomorphism & .803 & 25 & .036\\
    Animacy           & .593 & 26 & .054\\
    Likeability       & .118 & 26 & .063\\
    Perceived Intelligence & .487 & 27 & .049\\
    Perceived Safety       & .635 & 13 & .020\\ 
    \bottomrule
  \end{tabular}
  \label{tab:dependency}
\end{table}%
\autoref{tab:dependency} shows the $p$ values of the Fisher's exact test\footnote{The Fisher's exact test is implemented as \texttt{fisher.test()} in \texttt{R}'s in-built \texttt{stats} package.}. 
We see that the null can be rejected for Warmth and Competence only ($p<0.05$).
In other words, the participant's responses to these two dimensions are
dependent on the observed preferred condition reported directly by the participant, and vice versa.

Column $N$ of \autoref{tab:dependency} shows the number of considered participants.
Participants are considered when they reported a preference and when $e_d$ is defined
(i.e., there is exactly one condition with a maximum scale response value).

The uncertainty coefficient $U(o\mid{}e)$ quantifies the magnitude of above effect. It describes how consistent 
the expected preference $e$ can predict the observed condition $o$. 
The uncertainty coefficient $U$ measures the \emph{strength} between categorical association using the conditional entropy, i.e., the proportion of the reduced uncertainty~\cite{Nehmzow-06}. 
The uncertainty coefficient\footnote{It is also ambiguously referred to as Theil's $U$, a term which usually
refers to the $U$ statistics used in finance.} is commonly used to evaluate the effectiveness of cluster
algorithms. 
An interesting property is that it does not take into account any correspondence assumptions,
so it does not matter how the levels of $e$ and $o$ are hypothesized to be related.
This is a joint property with the Fisher's exact test, which makes $U$ a good choice for an effect size.
Note that $U$ is independent of the amount of levels of the variables (i.e., the size of the contingency table)
or the sample size of the study. This allows to compare the strength of association between this study and future studies.

$U$ is a directed effect. The interesting question for this study is: How much does the highest scale response to a dimension ($e_d$) tells us about the observed preferences ($o$), i.e., the participants' self-reported preference. More formally: What fraction of the remaining uncertainty of $o$ can be predicted given $e$:~$U(o|e)$. 
The results\footnote{$U$ is computed with \texttt{UncertCoef(table(o,e), direction=c("column"), p.zero.correction=T)} from the \texttt{R} package \texttt{DescTools}.} in \autoref{tab:dependency} show that $U$ is by far the highest for the dimensions Warmth and Competence---indicating that by itself, those two dimensions are by far the best predictors of self-reported human preference in our experiment. 

A value of $1$ would indicate that a given dimension reduces all remaining uncertainty in the prediction. The value of $U$ always lies between 0 and 1, which allows us to compare how much each dimension predicts the self-reported preference. We see that Warmth and Competence provide several times as much uncertainty reduction as the other dimensions.
Note that this is just the reduction of uncertainty by knowing which conditions had the maximal response for one singular dimension. If we would combine dimensions in \autoref{eq:expected_preferred}, or consider the scalar values\todo{mention this here?}, we could potentially achieve even better predictive power. 

\subsection{Correspondence Analysis}
The dependency analysis revealed that there is a statistically significant 
association between $e$ and $o$ for the dimensions Warmth and Competence. 
The question which remains is how the levels (i.e., conditions) of the two categorical variables $e$ and $o$ correspond to each other. For example, do participants who respond highest to the dimension Warmth in condition C ($e_{\text{Warmth}}=C$) also self-report to prefer this condition ($o=C$)?

\begin{figure}[!htbp]
  \begin{subfigure}{.5\textwidth}
    \centering
    \caption{Warmth}
    \includegraphics{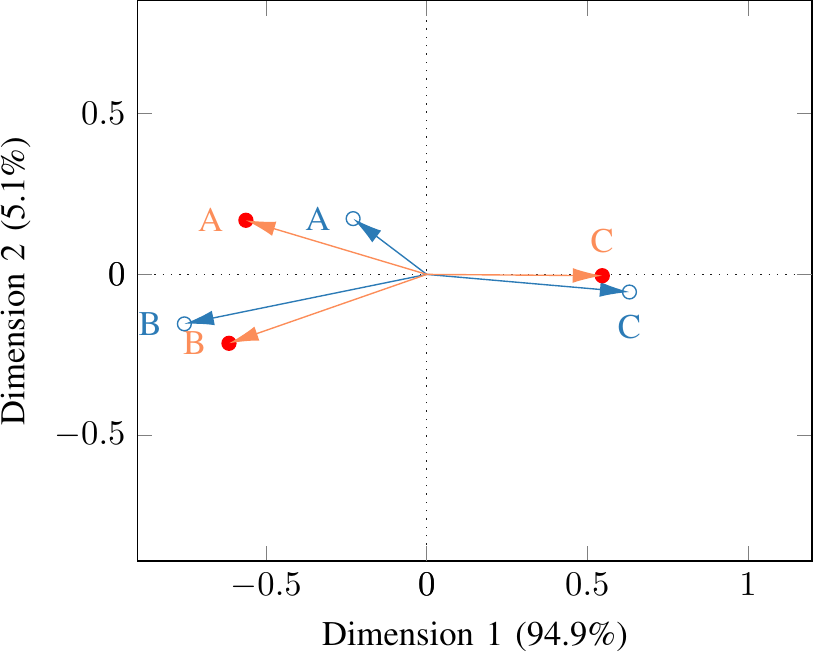}
    \label{fig:results:ca:Warmth}
  \end{subfigure}
  \begin{subfigure}{.5\textwidth}
    \centering
    \caption{Comeptence}
    \includegraphics{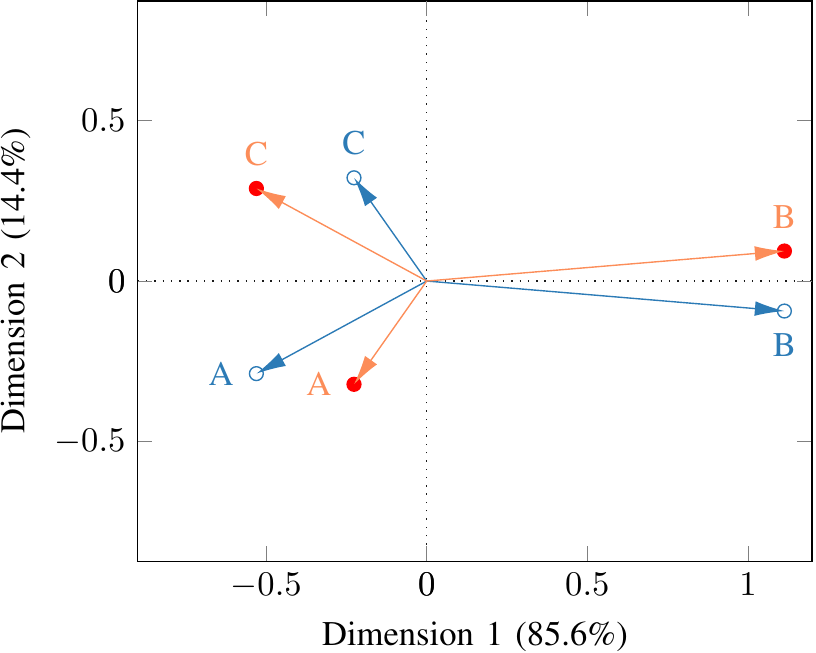}
    \label{fig:results:ca:Competence}
  \end{subfigure}
  \caption{The results of the correspondence analysis (CA) for the 
    levels of $o$ (\textcolor{color1}{blue}) and the levels of $e$ 
    (\textcolor{color2}{red}) for the dimensions 
    \subref{fig:results:ca:Warmth} Warmth and 
    \subref{fig:results:ca:Competence} Competence. For example, for the 
    dimension Warmth the condition C corresponds between $e$ and 
    $o$.} 
  \label{fig:results:ca}
\end{figure}

To answer this, we use correspondence analysis (CA), an extension of the Principal Component Analysis~(PCA) to categorical variables.
The analysis breaks down a higher dimensional table into fewer dimensions. This allows to plot the results and analyze the
correspondence graphically.
\autoref{fig:results:ca} shows the plotted results\footnote{Computed with \texttt{ca(table(o,e), arrows=c(T,T))} with the \texttt{R} package \texttt{ca}.} of the analysis for the predictors Warmth~\subref{fig:results:ca:Warmth} and Competence~\subref{fig:results:ca:Competence}.
A small angle between the arrows pointing from the coordinate origin to the levels of $o$~(\textcolor{color1}{blue}) and $e$~(\textcolor{color2}{red}) indicate a strong correspondence between these levels. An angle larger than \SI{90}{\degree} indicates no correspondence. The distance from the coordinate origin indicates the strength.

For Warmth, \autoref{fig:results:ca:Warmth} shows that the levels corresponding 
most to each other comprise the same conditions. For example, the level C of $e$ corresponds most to level C of $o$.
In other words, participants who report to prefer condition C respond highest to Warmth in C.
For the levels A and B the correspondence is not as strong, but also present. A from~$e$ is more associated to A from~$o$ and so it is with B.
This is remarkable, as the most computed condition of $e$ is C, which in turn only allows for fewer observations for A and B, which explains the \emph{weaker} correspondence. 
In contrast for the predictor dimension Competence~\subref{fig:results:ca:Competence}, the frequencies of $e$ are more evenly distributed, allowing for a more clear picture of the correspondence.
The angles between the level pairs are small, with the level~B being uniquely and most strongly associated.

It is safe to say that the statistically significant dependency is meaningful, i.e., the levels A, B and C of $e$ correspond to the levels A, B and C of $o$ respectively.
This is remarkable in the sense that we operate on a small sample size with quite small main effects as discussed earlier.

\section{DISCUSSION}
\label{sec:discussion}
The current study used the same robot platform, but different behavior generations. All of these different behaviors are hard to tell apart when observing~\cite{supp}. This is reflected in the results because there are only small differences in the dimensions between conditions and participants have no consensus for a preferred condition.
This is intentional and a central objective by the study design. It minimizes the chance of a common cause explanation, which would cause an arbitrary influence on the dimensions.
And yet, the dimensions Warmth and Competence are sensitive to the directly assessed participants' interaction preference.
This is remarkable and results presented above provide evidence that the dimensions Warmth and Competence are the best candidates for predicting participants preferred robot behavior in a human-robot interaction scenario.
The results suggest that they are the only candidates of all investigated dimensions.
This finding indicates that the two dimensions transferred to \ac{HRI} in the \ac{RoSAS} questionnaire can be used for interaction studies.

In~\cite{CarpinellaWymanEtAl-17} (study 2), it is discussed that the dimensions Warmth and Competence (RoSAS) are similar to the dimensions Likeability and Perceived Intelligence from the Godspeed questionnaire. %
To our surprise, we could not confirm any parallels between Competence and Perceived Intelligence.
Both dimensions show only very small effects among conditions~(cf.~\autoref{fig:main-effects}). The robot behavior does not, in fact, have any other goal than to explore. Any competence or intelligence rating would indeed surprise us.
And yet, although the little effect, Competence shows a clear dependency between the expected preferred interaction, and the observed participants interaction~(cf.~\autoref{tab:dependency}).
On the one hand, this dependency is strong evidence for Competence, as it is present despite the very small effects.
On the other hand, it is of surprise that the dimension Perceived Intelligence does not show a similarity.

More surprising is that Likeability fails at predicting the interaction preference. 
Considering that Likeability had almost the largest main effect for two condition pairs in \autoref{fig:main-effects} among all dimensions, this suggests that Likeability, contrary to its intuitive meaning, does not reveal much about the participants' robot behavior preference.
In addition, it seems that Godspeed's Likeability does not necessarily measure the same psychological construct as Warmth.

The dimension Discomfort shows statistical significance for the same condition pairs ($\text{A}-\text{B}$ and $\text{A}-\text{C}$) as Likeability (cf.~\autoref{fig:main-effects}).
In a sense, Discomfort\textsuperscript{--} seems to be the inverse dimension of Likeability.
And similar to Likeability, it does not reveal much information about the preferred interaction. Both observations are somewhat counter-intuitive. The dimensions are sufficiently sensitive and need to be considered carefully.

Overall our approach provides evidence that Warmth and Competence are the central dimensions for understanding human's preferred interaction. 

\section{LIMITATIONS and FUTURE WORK}
The proposed analysis in this paper can be easily incorporated into existing \ac{HRI} studies. For example, if an interaction study is planned already, adding a few questions at the end of the session for assessing participants' preference could provide further evidence for the strength of the dimensions Warmth and Competence as predictors for participant's preference.
That way, the \ac{HRI} community could learn from a variety of studies with different robot behaviors and robot platforms, and we could further understand if the dimensions Warmth and Competence are indeed applicable to \ac{HRI} in the same way as human-human interaction.
Ideally, the gathered knowledge brings us closer to a standardized measuring instrument for comparable \ac{HRI}.
We thus hope that other scientists adopt the underlying study design presented in \autoref{sec:study_design} for comparing the results with other robot platforms or with other behavior generators.

In future work we would like understand more about the ties of Warmth and Competence and the human perception of robots.
For example, if an interaction study is planned already, adding a few questions at the end of the session for assessing participants' perception of sympathy or pity, could provide further evidence for the strength of the dimensions Warmth and Competence.
It is known from social cognition that a pitied group, i.e., a group perceived as warm but incompetent, elicits helping behavior, but is neglected~\cite{CuddyFiskeEtAl-07}.
If more parallels are found, this would help in two ways: firstly, it allows to strengthen the ties between the fields of social cognition and \ac{HRI} and secondly, it would provide us with a tool to better predict behavior facilitation in \ac{HRI} scenarios.

One of the limitations of this study is that it is constrained to a specific robot platform. A more humanoid robot may reveal that Anthropomorphism from the Godspeed scale may be a good predictor for a human's interaction preference. This was partially found in the validation process of the \ac{RoSAS}~\cite{CarpinellaWymanEtAl-17}.
However, considering the strong evidence from social cognition that Warmth and Competence are the strongest indicators for almost all characterization and traits of humans~\cite{FiskeCuddyEtAl-07,JuddJames-HawkinsEtAl-05}, we would assume they will again play a strong role for interaction preferences.

\section{CONCLUSION}
\label{sec:conclusion}
In this study we assessed whether dimensions of the Godspeed questionnaire or the Robotic Social Attribute Scale~(RoSAS) can be used as a predictor for human's preference for interacting with a robot based on previous interactions with differently behaving robots.
The only discriminator among the conditions was the fairly similar, generated robot behavior. 
We found evidence that the central dimensions Warmth and Competence known from social cognition are the strongest predictors for participants' preference to interact again with a robot.
This indicates that, similar to inter persona interaction in social cognition, humans prefer to interact with robots perceived as more warm.

The proposed approach used for investigating the usability for the dimensions in an \ac{HRI} scenario can be easily extended to other \ac{HRI} studies, robots or interaction paradigms.


\end{document}